\documentclass[preprint2]{aastex}

\usepackage{graphicx}
\usepackage[authoryear]{}

\begin{document}

\title{The Occurrence of Non-pulsating Stars in the $\gamma$ Doradus/$\delta$ Scuti Pulsation Instability Region\footnote{Updated, corrected and LaTeX typeset version of  \citet{guzik13} paper published in Astronomical Review, Vol. 8.4, pp. 4-30, October 2013.}}

\author{J.A. Guzik\altaffilmark{a}, P.A. Bradley\altaffilmark{b}, J. Jackiewicz\altaffilmark{c}, K. Uytterhoeven\altaffilmark{d,e}, and K. Kinemuchi\altaffilmark{f}}

\altaffiltext{a}{X-Theoretical Design Division, Los Alamos National Laboratory, XTD-NTA, MS T-086, Los Alamos, NM  87545 USA  e-mail:joy@lanl.gov}
\altaffiltext{b}{X-Computational Physics Division, Los Alamos National Laboratory, XCP-6, MS F-699, Los Alamos, NM  87545 USA}
\altaffiltext{c}{Department of Astronomy, New Mexico State University, Las Cruces, NM  88003  USA}
\altaffiltext{d}{Instituto de Astrof'sica de Canarias (IAC), E-38200 La Laguna, Tenerife, Spain}
\altaffiltext{e}{Universidad de La Laguna, Dept. Astrof{\`i}sica, E-38206 La Laguna, Tenerife, Spain}
\altaffiltext{f}{Apache Point Observatory, Sunspot, NM 88349  USA}

\begin{abstract}
As part of the NASA {\it Kepler} Guest Observer program, we requested and obtained long-cadence data on about 2700 faint ({\it Kepler} magnitude 14-16) stars with effective temperatures and surface gravities that lie near or within the pulsation instability region for main-sequence $\gamma$ Doradus and $\delta$ Scuti pulsating variables. These variables are of spectral type A-F with masses of 1.4 to 2.5 solar masses.  The $\delta$ Scuti stars pulsate in radial and non-radial acoustic modes, with periods of a few hours (frequencies $\sim$10 cycles/day), while $\gamma$ Doradus variables pulsate in nonradial gravity modes with periods 0.3 to 3 days (frequencies $\sim$1 cycle/day).   Here we consider the light curves and Fourier transforms of 633 stars in an unbiased sample observed by {\it Kepler} in Quarters 6-13 (June 2010-June 2012).  We show the location of these stars in the log surface gravity--effective temperature diagram compared to the instability region limits established from ground-based observations, and taking into account uncertainties and biases in the {\it Kepler} Input Catalog T$_{\rm eff}$ values.  While hundreds of variables have been discovered in the {\it Kepler} data, about 60\% of the stars in our sample do not show any frequencies between 0.2 and 24.4 cycles/day with amplitude above 20 parts per million.  We find that six of these apparently constant stars lie within the pulsation instability region. We discuss some possible reasons that these stars do not show photometric variability in the {\it Kepler} data.  We also comment on the `non-constant' stars, and on 26 variable-star candidates, many of which also do not lie within the expected instability regions.

\end{abstract}


\section{Introduction}
\label{intro}

The NASA {\it Kepler} spacecraft was launched March 7, 2009 with the mission to detect transits of Earth-sized planets around sun-like stars using high-precision CCD photometry \citep{borucki10}.  As a secondary mission, {\it Kepler} photometry has also been used to survey many thousands of stars for variability and stellar activity \citep{gilliland10}.  The {\it Kepler} field of view is fixed in the Cygnus-Lyra region and is about 115 square degrees on the sky.

The $\gamma$ Doradus stars are late-A to F spectral-type main-sequence variables showing multiple periods of 0.3 to 3 days that have been identified as high-order, low-degree nonradial gravity (g) modes.  The intrinsic variability of the prototype $\gamma$ Dor was discussed by \citet{balona94}, and the new class with 15 members was described by \citet{kaye99}.  Since then, over 60 members of the class have been identified \citep{henry07} using ground-based observations.  The $\delta$ Scuti stars, on the other hand, pulsate in low-order radial and nonradial pressure (p) modes, although some of their modes have mixed p- and g-mode character; about 630 were catalogued as of 2000 \citep{rodriguez00}.  Their instability strip lies on the main sequence between early A and early F, and their periods range from just under 30 minutes to about 0.3 days.  Their pulsations are driven by the ionization of helium increasing the opacity and regulating radiation flow in the envelope layers at 50,000 K \citep[the `$\kappa$ effect'; see, e.g.,][]{chevalier71}.

The $\gamma$ Dor and $\delta$ Sct stars are interesting asteroseismically because they form a bridge between the lower-mass solar-like stars with large convective envelopes, long lifetimes, and slow rotation, and the short-lived, massive, rapidly rotating stars with convective cores. The $\delta$ Sct and $\gamma$ Dor stars have both convective cores and convective envelopes, a variety of rotation rates, and live long enough for element settling, diffusion, and radiative levitation to alter their surface abundances and cause chemical peculiarities.  The envelope convection zones of the hotter and more massive $\delta$ Sct stars are confined to smaller regions around 10,000-50,000 K where hydrogen and helium are ionizing, and where convection is transporting a small fraction of the luminosity; the envelope convection zones become deeper and carry a larger fraction of the star's luminosity with increasing stellar age or decreasing stellar mass \citep[see, e.g.,][and references therein]{turcotte98}.  Both $\gamma$ Dor and $\delta$ Sct stars pulsate in multiple modes that probe both the core and envelope.  The pulsations of these stars can be used to diagnose these phenomena and test stellar models.

After their discovery, the theoretical pulsation mechanism for $\gamma$ Dor stars was pursued.  Early thoughts were that settling/levitation might concentrate iron into a region of the envelope where a $\kappa$-effect like mechanism might operate \citep[see, e.g.,][]{guzik00}, similar to the now-accepted pulsation mechanism for driving g modes in subdwarf B stars.  A second idea was that `convective driving' at the top of an envelope convection zone might drive g modes, analogous to the way they are driven in white-dwarf pulsators \citep{wugoldreich00}.  \citet{guzik00} proposed the mechanism of `convective blocking' at the base of the envelope convective zones of $\gamma$ Dor stars.  When the temperature at the convective envelope base is between 200,000 and 500,000 K, the convective timescale (mixing length/convective velocity) at this location is comparable to or longer than the g-mode pulsation period.  Since convection takes a portion of the pulsation cycle to turn on and transport the emergent luminosity, luminosity is periodically blocked, resulting in pulsation driving. This mechanism predicted unstable g modes in exactly the observed period range, and has become the accepted mechanism for $\gamma$ Dor pulsation driving \citep[see also][]{dupret04,dupret05,grigahcene06}.  Finally, it has been proposed that some frequencies might be stochastically excited as in solar-like oscillations \citep{pereira07, houdek99, samadi02, antoci11}.

Hybrid $\gamma$ Dor/$\delta$ Sct stars are among the most interesting targets for asteroseismology because the two types of modes (pressure and gravity) probe different regions of the star and are sensitive to the details of the two different driving mechanisms.  Because these driving mechanisms are somewhat mutually exclusive, hybrid stars exhibiting both types of pulsations are expected to exist only in a small overlapping region of temperature-luminosity space in the Hertzsprung-Russell (H-R) diagram.

Before the advent of the {\it Kepler} and CoRoT space missions, only four hybrid $\gamma$ Dor/$\delta$ Sct pulsators had been discovered. However, the first analysis by the {\it Kepler} Asteroseismic Science Consortium (KASC) of 234 or targets showing pulsations of either type revealed hybrid behavior in essentially all of them \citep{grigahcene10}.  In a study of 750 KASC A-F stars observed for four quarters \citep{uytterhoeven11}, 475 stars showed either $\delta$ Sct or $\gamma$ Dor variability, and 36\% of these were hybrids.  The {\it Kepler} hybrids are not confined to a small overlapping region of the two instability types in the temperature-luminosity space of the H-R diagram as predicted by theory.  Instead, they are observed throughout both the predicted $\gamma$ Dor and $\delta$ Sct instability regions, and even at cooler and hotter temperatures outside these regions.  Despite extensive study of this large sample and of the public data \citep{balonadziembowski11}, no obvious frequency or amplitude correlations with stellar properties have emerged, and there seems to be no clear separation of $\gamma$ Dor and $\delta$ Sct pulsators in the H-R diagram.  The known driving mechanisms cannot explain the pulsation behavior.

The existence and properties of these {\it Kepler} variable star candidates raise a number of questions:  Why are hybrids much more common than predicted by theory?  Are additional pulsation driving mechanisms at work?  Is {\it Kepler} detecting modes of high degree that are usually not visible in ground-based photometric data?  Why are there apparently `constant' stars that lie within the instability regions but show no pulsation frequencies in the $\gamma$ Dor or $\delta$ Sct frequency range?

Through the NASA {\it Kepler} Guest Observer program (Cycles 1-4), we requested and received observations of about 2700 stars of spectral types A-F.  Our goals are to search for pulsating $\gamma$ Dor and $\delta$ Sct variable star candidates, identify ÔhybridÕ stars pulsating in both types that are especially useful for asteroseismology and testing stellar pulsation theory, characterize the frequency content, and look for amplitude variability.  In the course of these observations, we discovered many eclipsing-binary stars \citep[see][]{gaulme14}, stars that are probably not pulsating but show low-frequency photometric variations due to stellar activity or rotating star spots, and many stars that show no variability at the frequencies expected for $\gamma$ Dor or $\delta$ Sct stars.  This paper focuses on these apparently `constant' stars in the sample from our Cycle 2 and 3 Guest Observer data, and discusses whether they should be expected to pulsate.  We also consider the stars in the sample that appear unambiguously to be $\gamma$ Dor or $\delta$ Sct candidates, and discuss their position in the log g--T$_{\rm eff}$ diagram relative to the $\gamma$ Dor or $\delta$ Sct instability regions established from ground-based observations.  

\section{Target Selection and Data Processing}

We examined the {\it Kepler} data on the sample of 633 stars requested for {\it Kepler} Guest Observer Cycles 2 and 3, taken in Quarters 6 through 13, spanning the time period June 24, 2010 through June 26, 2012.  These stars are relatively faint, with {\it Kepler} magnitudes 14-15.5.   We requested only long-cadence data, with an integration time per data point of 29.4 minutes.  The Fourier transform of this data will only be able to detect frequencies less than the Nyquist frequency, or half the sampling frequency, which is 24.4 cycles/day.  However, since $\gamma$ Dor stars have frequencies of about 1 cycle/day and most $\delta$ Sct stars have frequencies of 10-20 cycles/day, the long-cadence data are adequate to identify $\gamma$ Dor and most $\delta$ Sct candidates.

We chose these target stars to observe by using the {\it Kepler} Guest Observer Target selection tool to search the {\it Kepler} Input Catalog \citep[KIC,][]{latham05}.  We restricted the sample to targets that were likely to be in or near the $\gamma$ Dor/$\delta$ Sct instability strips, with effective temperature 6200 $<$ T$_{\rm eff}$ $<$ 8300 K; log surface gravity 3.6 $<$ log g $<$ 4.7 contamination or blend from background stars $<$ 10$^{-2}$, and {\it Kepler} Flag 0, meaning that these targets had not yet been observed by {\it Kepler}.  Most of the brighter stars had already been observed by the {\it Kepler} Science Team or KASC, and these are discussed by \citet{uytterhoeven11}.  We have also received observations for Quarter 14-16 from our Cycle 4 Guest Observer proposal on over 2000 additional stars;  however, the target selection for Cycle 4 was not unbiased, but instead was cross-correlated with a list of potentially variable stars identified by comparing full-frame images taken during the {\it Kepler} spacecraft commissioning period.  These stars will be treated in a separate paper.  We also did not include the small sample of 14 stars observed for our Cycle 1 GO proposal, because these stars were either chosen from pre-{\it Kepler} lists of known variables, or were restricted to a very specific T$_{\rm eff}$ range to find hybrids.

After downloading the data files from the MAST (Mikulski Archive for Space Telescopes, http://archive.stsci.edu/index.html), we processed them using Matlab scripts developed by J. Jackiewicz.  We combined data for all available quarters for each star.  We chose to use the data corrected by the {\it Kepler} pipeline; we have also tried using the raw light curves, and found essentially no difference in the frequencies found by the Fourier transform.  

The Matlab scripts remove outlier points and interpolate the light curves to an equidistant time grid.  The light curves are then converted from {\it Kepler} flux (F$_K$) to parts per million (ppm) using the formula f(t) = 10$^6$(F$_K$/y - 1), where y is either the mean value of the entire light curve, or a low-order polynomial fit to the light curve, depending on artifacts present in the data.  The fitting did well in removing long-term trends that are non-physical in origin \citep[see also][]{mcnamara12}.  After the processing and Fourier transform of the data, the Matlab scripts generate a plot for each star, such as the examples shown later in Figs. \ref{4731085} through \ref{10910954}.  In these plots, the bottom panel is the light curve in ppm vs. time; the top panel shows the amplitude in ppm vs. frequency in cycles/day determined by the Fourier transform, and the middle two panels are enlargements of the frequency ranges 0-5 cycles/day and 5-24.4 cycles/day.  The header on each panel shows the KIC number, the Quarters of observation, the {\it Kepler} magnitude, and log surface gravity (log g).  The radius in the header is also from the KIC but is not accurate, as it is derived by assuming a stellar mass of 1 M$_{\odot}$  and using the log surface gravity (log g) from the KIC; also in this header, the log g is rounded to two digits.

\section{Identification and Distribution of `Constant' Stars in the H-R Diagram}

\begin{figure*}%tb
\center
\includegraphics[width=1.5\columnwidth]{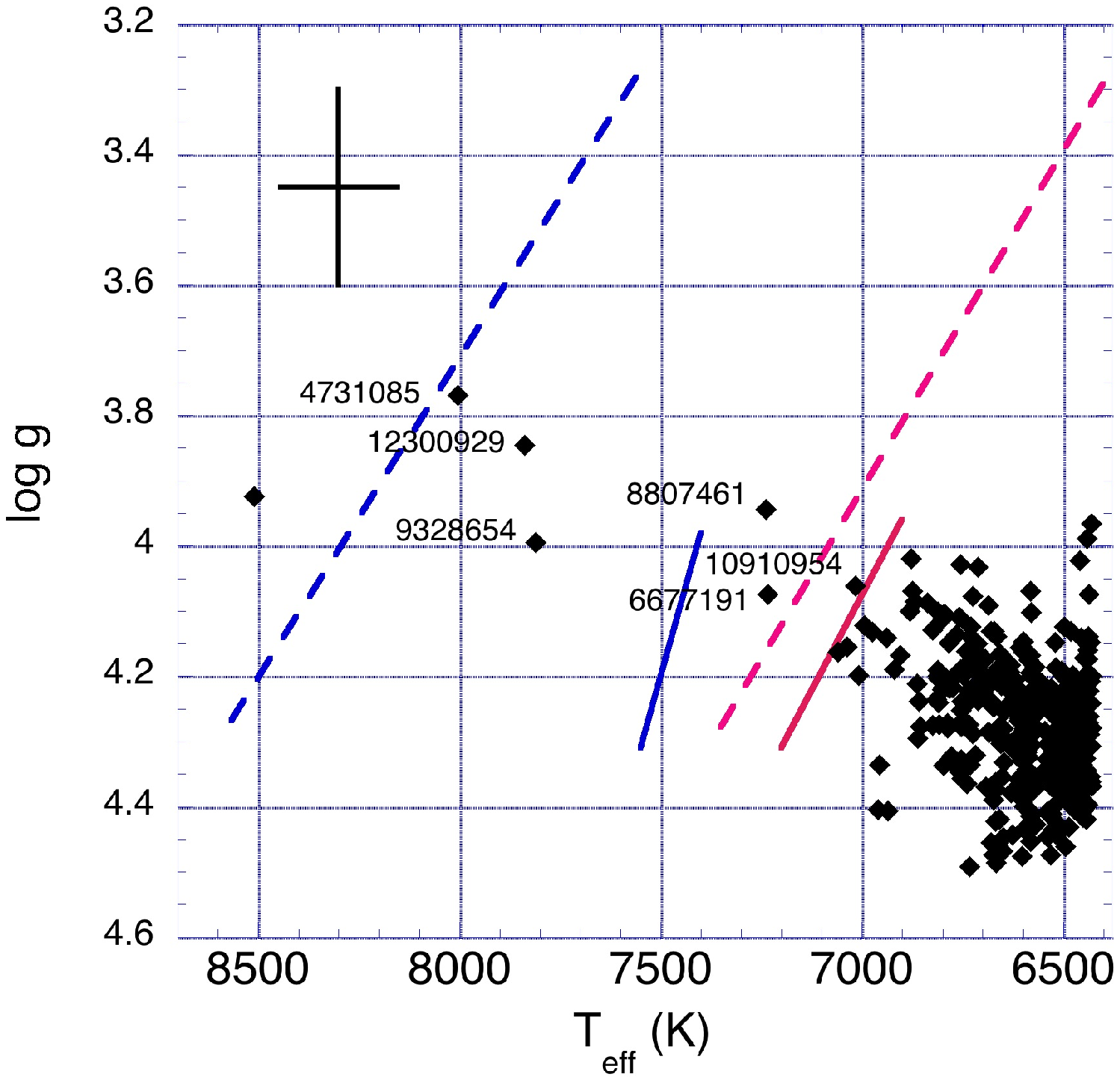}
\caption{Location of sample stars that are `constant' (see definition in text) in the log surface gravity -- T$_{\rm eff}$ diagram, along with $\delta$ Sct (dashed lines) and $\gamma$ Dor (solid lines) instability-strip boundaries established from pre-{\it Kepler} ground-based observations \citep{rodriguezbreger01, handlershobbrook02}.  The T$_{\rm eff}$ of the sample stars has been shifted by +229 K to account for the systematic offset in stellar effective temperatures between the {\it Kepler} Input Catalog and SDSS photometry for this temperature range, as determined by \citet{pinsonneault12}.  The black cross shows an error bar on log g (0.3 dex) and T$_{\rm eff}$ (290 K) established by comparisons of KIC values and values derived from ground-based spectroscopy for brighter {\it Kepler} targets \citep{uytterhoeven11}.}
\label{ConstantStarLargerSym}
\end{figure*}

\begin{figure*}%tb
\center
\includegraphics[width=1.5\columnwidth]{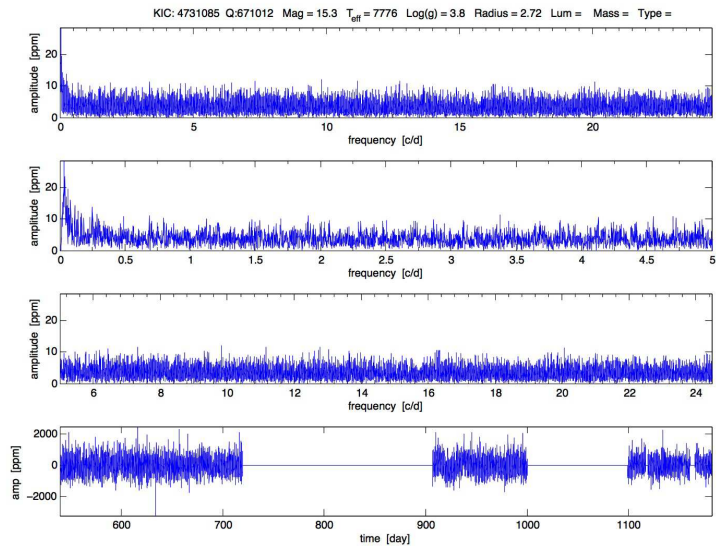}
\caption{Fourier transform (top 3 panels) and light curve (bottom panel) for KIC 4731085, the hottest of the `constant' stars within the $\delta$ Sct instability-strip boundary in Fig. \ref{ConstantStarLargerSym}.  The noise level of this faint star is at about the 10 ppm level.  Some frequencies may be present at $\sim$10 ppm amplitude, but are not detectable with this data set.}
\label{4731085}
\end{figure*}

\begin{figure*}%tb
\center
\includegraphics[width=1.5\columnwidth]{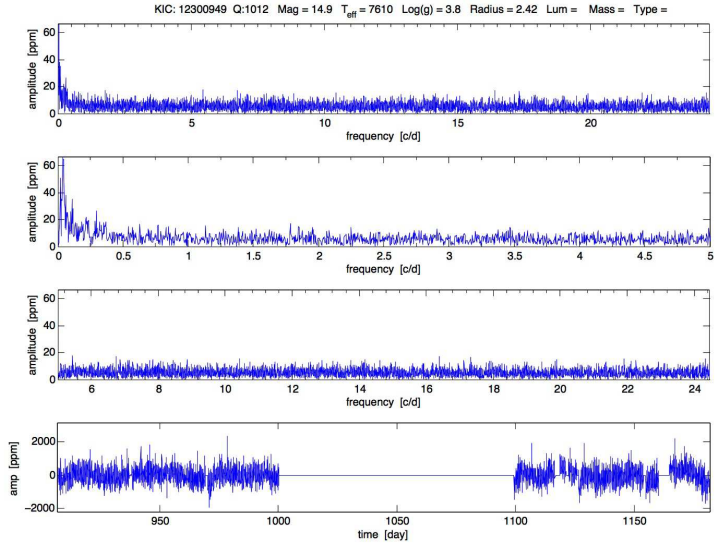}
\caption{Fourier transform (top 3 panels) and light curve (bottom panel) for KIC 12300949, another hot `constant' star within the $\delta$ Sct instability region in Fig. \ref{ConstantStarLargerSym}.  The noise level is about 15 ppm; some frequencies with amplitude $\sim$20 ppm may be present that could be detected with a longer time series.}
\label{12300949}
\end{figure*}

\begin{figure*}%tb
\center
\includegraphics[width=1.5\columnwidth]{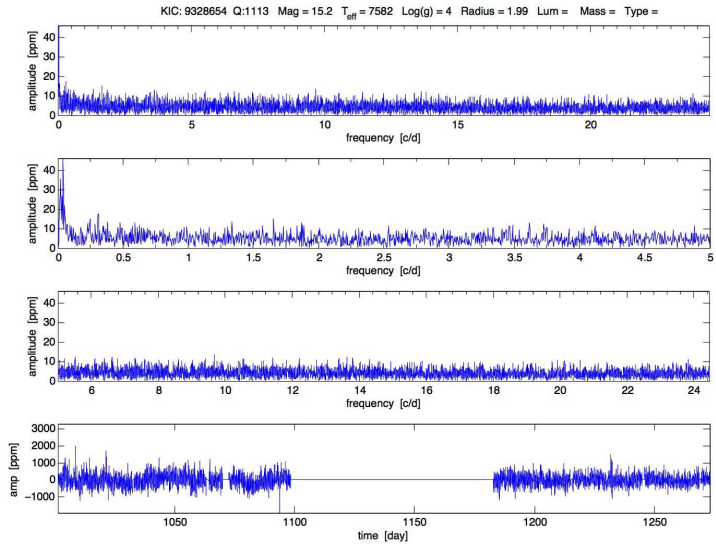}
\caption{Fourier transform (top 3 panels) and light curve (bottom panel) for KIC 9328654, another hot `constant' star within the $\delta$ Sct instability-strip boundary in Fig. 1.  The noise level is at about 5-10 ppm; some frequencies may be present with amplitude $\sim$10 ppm that could be detected with a longer time series.}
\label{9328654}
\end{figure*}

\begin{figure*}%tb
\center
\includegraphics[width=1.5\columnwidth]{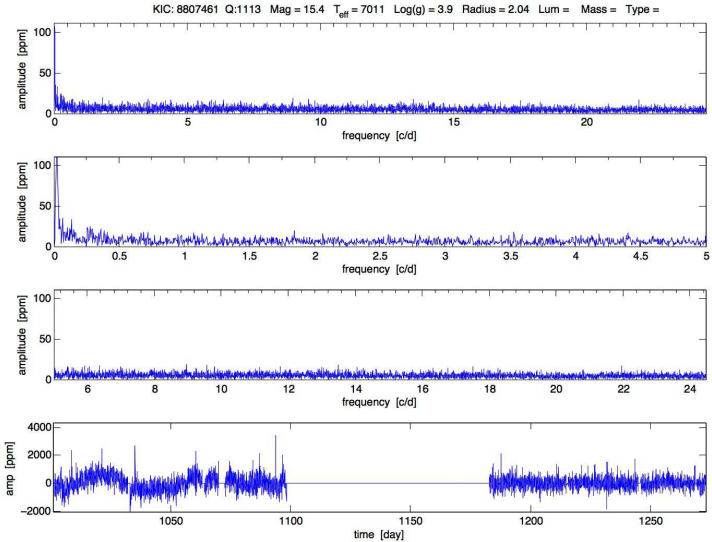}
\caption{Fourier transform (top 3 panels) and light curve (bottom panel) for KIC 8807461, a `constant' star that lies within the instability region where ÔhybridÕ $\gamma$ Dor-$\delta$ Sct stars are expected in Fig. \ref{ConstantStarLargerSym}.  The noise level is extremely low for such a faint star, and no frequencies are evident down to the few-ppm level.}
\label{8807461}
\end{figure*}

\begin{figure*}%tb
\center
\includegraphics[width=1.5\columnwidth]{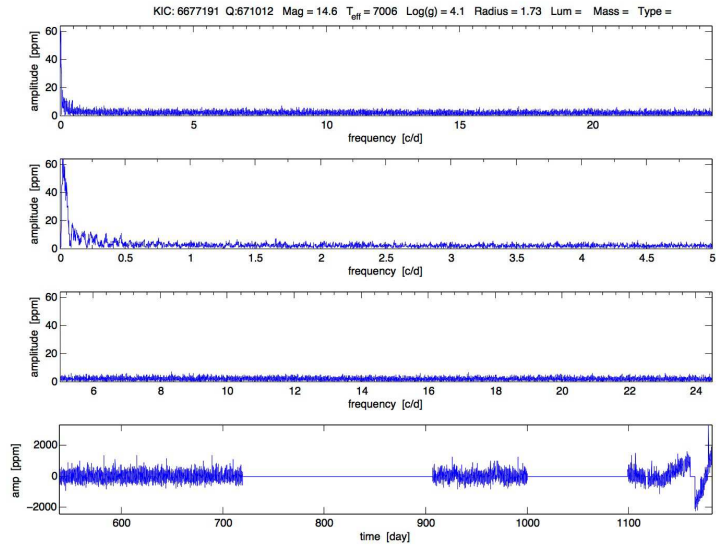}
\caption{Fourier transform (top 3 panels) and light curve (bottom panel) for KIC 6677191, a `constant' star that lies within the instability region where ÔhybridÕ $\gamma$ Dor-$\delta$ Sct stars are expected in Fig. \ref{ConstantStarLargerSym}.  A glitch in the signal level is evident in Quarter 12 of the light curve that was not removed by the {\it Kepler} data-reduction pipeline or polynomial-fit processing.   The noise level is very low, and no frequencies are evident in the $\gamma$ Dor or $\delta$ Sct frequency range (above 0.5 c/d) down to the few-ppm level.}
\label{6677191}
\end{figure*}

\begin{figure*}%tb
\center
\includegraphics[width=1.5\columnwidth]{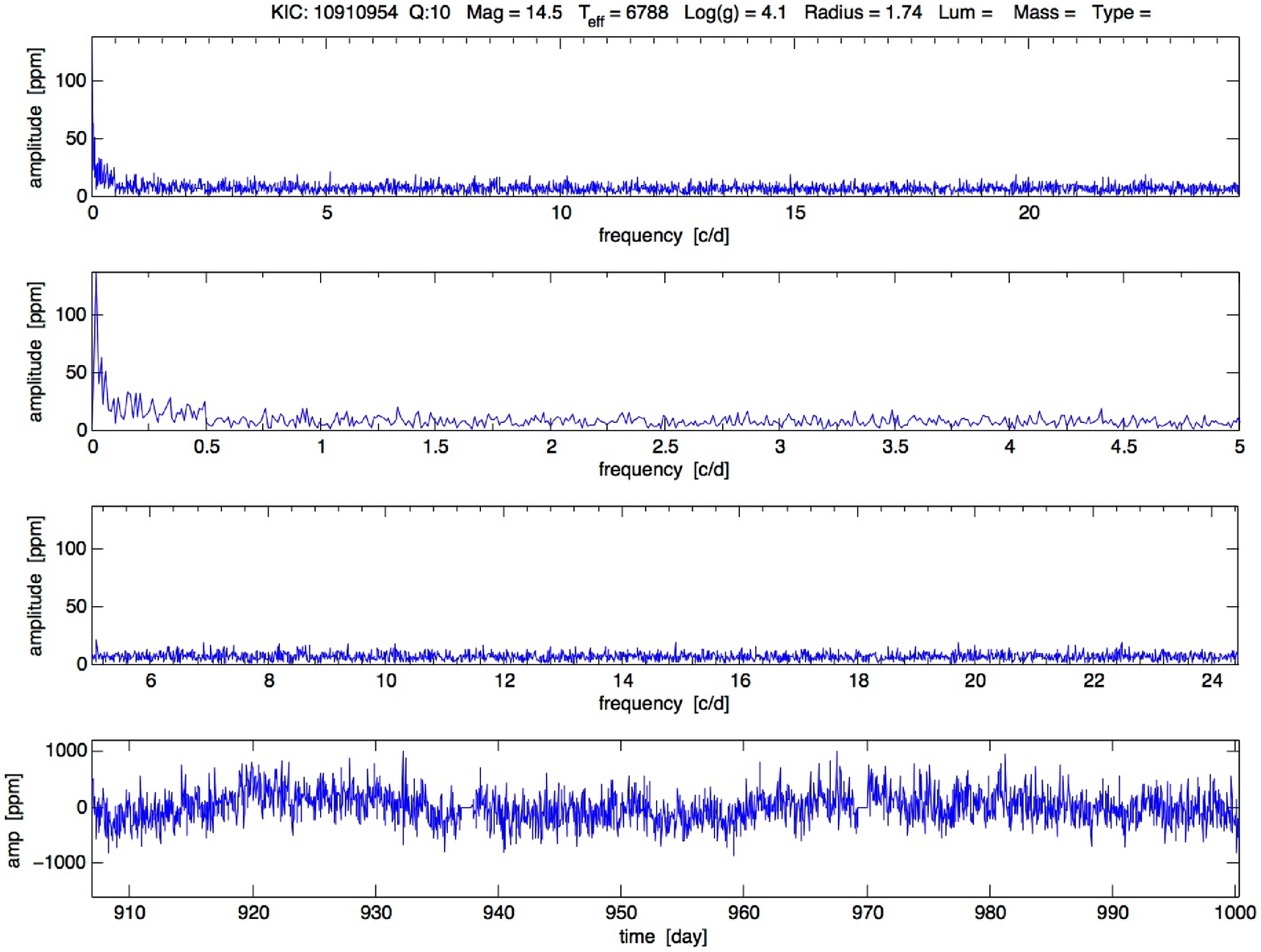}
\caption{Fourier transform (top 3 panels) and light curve (bottom panel) for KIC 10910954, a `constant' star that lies within the $\gamma$ Dor instability region of Fig.  \ref{ConstantStarLargerSym}. No frequencies are evident in the $\gamma$ Dor or $\delta$ Sct frequency range (above 0.5 c/d) down to the few-ppm level.  Only one Quarter (3 months) of data is available for this star, so it is possible that a longer time series would reveal some very low-amplitude modes.}
\label{10910954}
\end{figure*}

We examined individually the Matlab-generated plots for our sample of 633 stars.  We flagged as `constant' the stars with no power in the spectrum above 20 ppm between 0.2 to 24.4 c/d.

There is some judgment involved in deciding whether a given star meets these criteria.  For some stars, the noise level over a good portion of this frequency range is about 20 ppm.  For many stars, stellar activity may cause photometric variations and power in the spectrum at the 20 ppm level or higher at 0.2 to 0.5 c/d frequencies.  We eliminated stars that appear to be eclipsing binaries that may have stellar components that individually turn out to be either `constant' or variable once the binary signal is removed.

By these criteria, we find 359 `constant' stars out of the 633 stars in our sample, or roughly 60\%. 

For all of the stars in our sample, we applied a 229 K increase to the effective temperature given by the KIC. This shift is used to roughly account for the systematic offset between temperatures given in the KIC and Sloan Digital Sky Survey photometry for the temperature range of our stellar sample as determined by \citet{pinsonneault12}.  To determine this shift, we averaged the five temperature offsets given in Table 8 of Pinsonneault et al. for effective temperatures 6200-6600~K.  The 1-sigma error in these offsets is about 35~K.  

We next plotted the position of the `constant' stars in a log surface gravity vs.~T$_{\rm eff}$  diagram (Fig. 1), along with the $\delta$ Sct (dashed lines) and $\gamma$ Dor (solid lines) instability-strip boundaries established from pre-{\it Kepler} ground-based observations \citep{rodriguezbreger01, handlershobbrook02}.   We also show with a black cross the error bar on log g (0.3 dex) and T$_{\rm eff}$ (290 K) derived by comparing KIC values of brighter {\it Kepler} targets with those derived from ground-based spectroscopy, as discussed by \citet{uytterhoeven11}.

The uncertainties on T$_{\rm eff}$ and log g for our sample may mean that some of the stars in our sample may move out of the instability regions, but, just as easily, some of the stars just outside the instability region may move into the instability region.  If the uncertainties are random and not systematic, our conclusion is that of order half a dozen, or about 2\% of this sample is within these instability-strip boundaries.

Note that the instability-strip boundaries in Fig. 1 were derived from ground-based pre-{\it Kepler} observations of known $\gamma$ Dor and $\delta$ Sct stars with relatively accurate determinations of T$_{\rm eff}$ and log g from multicolor photometry or spectroscopy.  As discussed in Section IV, the instability region boundaries derived from theoretical stellar evolution models depend on many factors, including helium and element abundance, abundance mixtures, and convection treatment.   The instability region boundaries predicted by theory were in relatively good agreement with the ground-based observations prior to the {\it Kepler} data; however, the distribution of the many new $\delta$ Sct and $\gamma$ Dor candidates studied by KASC indicate that the instability-strip boundaries may be wider than predicted by theoretical models \citep[see, e.g.,][]{uytterhoeven11}, or that additional pulsation instability mechanisms may need to be considered to explain the data.

Figure  \ref{ConstantStarLargerSym} shows that only six of the 359 `constantÕ' stars lie within the instability-strip boundaries derived from ground-based variable star observations. Taking into account uncertainties in log g or T$_{\rm eff}$, several stars in the sample may move into or outside of the instability region boundaries.

We next examine the light curves and Fourier transforms of these six stars in more detail (Figs. \ref{4731085}, \ref{12300949}, \ref{9328654}, \ref{8807461}, \ref{6677191} and \ref{10910954}) in order of their effective temperature.  The figure captions include comments on the light curve and Fourier transform details.  It is interesting that the noise level varies among these stars; a few show no frequencies down to the 5 ppm level; a few show `noise' at the 10-20 ppm level throughout the frequency range, and some hint that a few frequencies may be present just above the noise level.

Most of the `constant' stars show some signal below 0.5 c/d, and sometimes a signal above 20 ppm at 0-0.1 c/d; as discussed above, we omitted this low frequency region from consideration in selecting `constant' stars.

\section{Explanations for `Constant' Stars}

Even though there are only a few stars in our sample that are `constant' but should be pulsating, these stars are interesting and should be explained.  Among the possible explanations that should be explored are:

1)  The stars are pulsating at $\delta$ Sct frequencies higher than 24.4 c/d. We have not considered this frequency region, as it is higher than the Nyquist frequency of our data set.  Short-cadence {\it Kepler} data ($\sim$1 minute sampling rate) generally is not available for such faint stars.  Many of these stars could even be higher-frequency solar-like oscillators.  Solar-like oscillations were detected using short-cadence {\it Kepler} data for the bright main-sequence F star $\theta$ Cyg, with T$_{\rm eff}$ $\sim$6700 K \citep[see, e.g.,][]{guzik11}.

2) The stars may be pulsating in higher spherical harmonic degree modes (e.g., degree$>$0-3) that aren't easily detectable photometrically. Such high-degree modes may be detectable spectroscopically, but high resolution and time series spectra for such faint stars requires continuous observation for weeks with 10-meter class telescopes, and is impossible to organize.  In addition, it is thought that the photometric variations may not completely average out for modes with degree up to 10-20 and may be visible with space-based photometry \citep[see][]{daszynska06,balonadziembowski99,barban01}.  High-degree modes (degree up to 14) that were identified using spectroscopic line profile variation analysis have been detected photometrically in the CoRoT light curve for the $\delta$ Sct star HD 50844 \citep{poretti09}. 

3) As discussed in the figure captions, there may be pulsation modes with amplitudes at or below the noise level of this data; the modes could possibly be detected with a longer time series or by reducing the noise levels, possibly making use of the {\it Kepler} pixel data.

4)  It is possible that a physical mechanism is operating that inhibits pulsations for some stars.  For example, diffusive helium settling might turn off $\delta$ Sct pulsations, or diffusion of metals in $\gamma$ Dor stars may cause the convection zone to become too shallow to enable the convective blocking mechanism for pulsation driving. Diffusive settling is predicted to deplete most of the helium and heavier elements from the photospheres of A-F stars during their main-sequence lifetimes.  Since elements are observed in the photospheres of most A-F stars (although some have peculiar abundance distributions), it is probable that some mechanism such as turbulence, or rotational mixing is inhibiting diffusive settling for most stars.  More work is necessary to determine how much diffusion is needed to inhibit pulsations, and to further investigate the effects of turbulence, rotation and magnetic activity on pulsation driving. 

5) The log g or T$_{\rm eff}$ in these `constant' stars may simply be in error, so that the stars are in reality outside the pulsation instability regions.  If the observational errors are random, some of the stars in our sample may move out of the instability regions, but, just as easily, some of the stars slightly outside the instability region may move into the instability region.  For our faint stars, spectroscopic observations to determine log g and T$_{\rm eff}$, and in addition the stellar metallicity, to high enough precision to settle this question are very expensive in telescope time, and may not even be possible.  A systematic effort is underway to derive more accurate effective temperatures, surface gravities, and element abundances for KASC target stars \citep{uytterhoeven10} that will better inform the uncertainties on these quantities for our fainter sample stars.

A possibly more important question/problem that has been motivated by the {\it Kepler} data is the presence of $\gamma$ Dor and $\delta$ Sct stars or hybrids well outside of their expected instability regions determined by either ground-based observations or stellar modeling \citep[see, e.g.,][]{uytterhoeven11,grigahcene10}.  Additional pulsation driving mechanisms may be needed to explain the observed frequencies.  Theoretically-predicted pulsation instability regions depend on stellar abundances, mixing-length treatment, mixing and diffusive settling, the treatment of pulsation driving and damping mechanisms, turbulent pressure, rotation, and probably other factors, so much more theoretical work is needed.

\section{Distribution of `Non-Constant' Stars}

\begin{figure*}%tb
\center
\includegraphics[width=1.5\columnwidth]{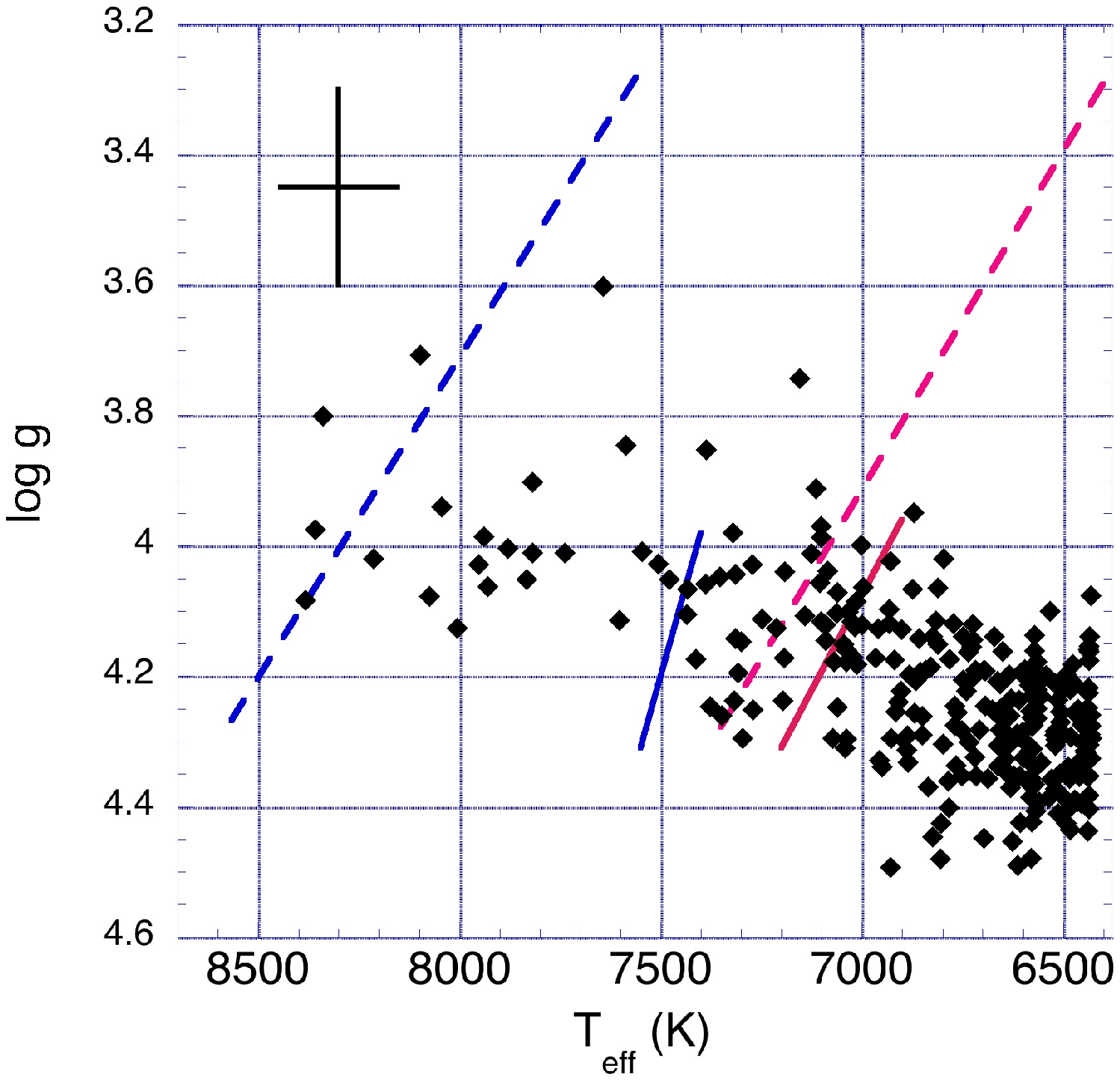}
\caption{Location of sample stars that are not `constant' (see definition in text) in the log g --T$_{\rm eff}$ diagram, along with the $\delta$ Sct (dashed lines) and $\gamma$ Dor (solid lines) instability-strip boundaries established from pre-{\it Kepler} ground-based observations \citep{rodriguezbreger01,handlershobbrook02}.  The T$_{\rm eff}$ of the stars has been shifted by +229 K to account for the systematic offset of stellar effective temperatures between the {\it Kepler} Input Catalog and SDSS photometry for this temperature range, as determined by \citet{pinsonneault12}.  The black cross shows an error bar on log g (0.3 dex) and T$_{\rm eff}$ (290 K) established by comparisons of KIC values and values derived from ground-based spectroscopy for brighter {\it Kepler} targets \citep{uytterhoeven11}. Most of these stars have light curves and Fourier transforms consistent with stellar activity (spots) or binarity.}
\label{NonconstantStarLargerSym}
\end{figure*}

\begin{figure*}%tb
\center
\includegraphics[width=1.5\columnwidth]{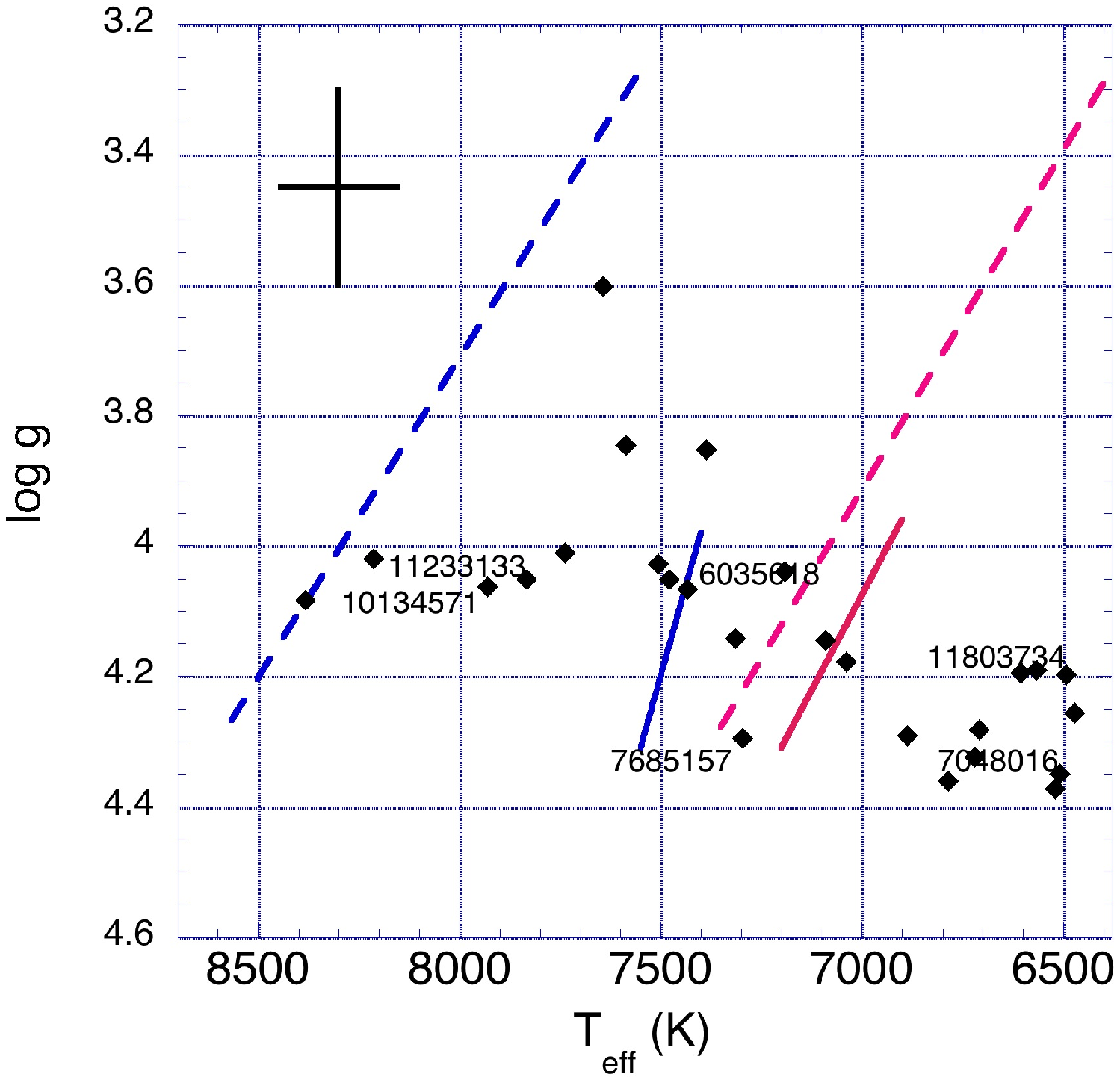}
\caption{Location of 26 stars that are $\gamma$ Dor or $\delta$ Sct candidates based on light curve and Fourier transform, in log g--T$_{\rm eff}$ diagram, along with $\delta$ Sct (dashed lines) and $\gamma$ Dor (solid red and blue lines) instability-strip boundaries established from pre-{\it Kepler} ground-based observations \citep{rodriguezbreger01,handlershobbrook02}.  The T$_{\rm eff}$ of the stars has been shifted by +229 K to account for the systematic offset of stellar effective temperatures between the {\it Kepler} Input Catalog and SDSS photometry for this temperature range, as determined by \citet{pinsonneault12}.  The black cross shows an error bar on log g (0.3 dex) and T$_{\rm eff}$ (290 K) established by comparisons of KIC values and values derived from ground-based spectroscopy for brighter {\it Kepler} targets \citep{uytterhoeven11}.  Some of these stars are cooler than the $\gamma$ Dor instability strip red edge, even accounting for uncertainties in T$_{\rm eff}$ and log g.  Many $\gamma$ Dor and $\delta$ Sct candidates were also found outside of the instability region boundaries in the KASC sample of generally brighter {\it Kepler} stars by \cite{uytterhoeven11}.}
\label{gammaDordeltaScutiLargerSym}
\end{figure*}

\begin{figure*}%tb
\center
\includegraphics[width=1.5\columnwidth]{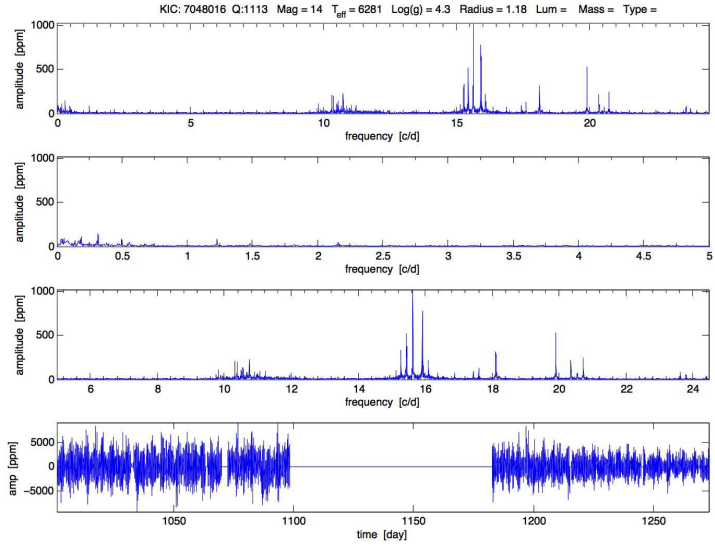}
\caption{KIC 7048016 light curve (bottom panel) and Fourier transform showing many frequencies in the $\delta$ Sct range.  However, this star has T$_{\rm eff}$ = 6510 K (including the temperature offset from the KIC value) and lies redward of the expected pulsation instability region.  }
\label{7048016}
\end{figure*}

\begin{figure*}%tb
\center
\includegraphics[width=1.5\columnwidth]{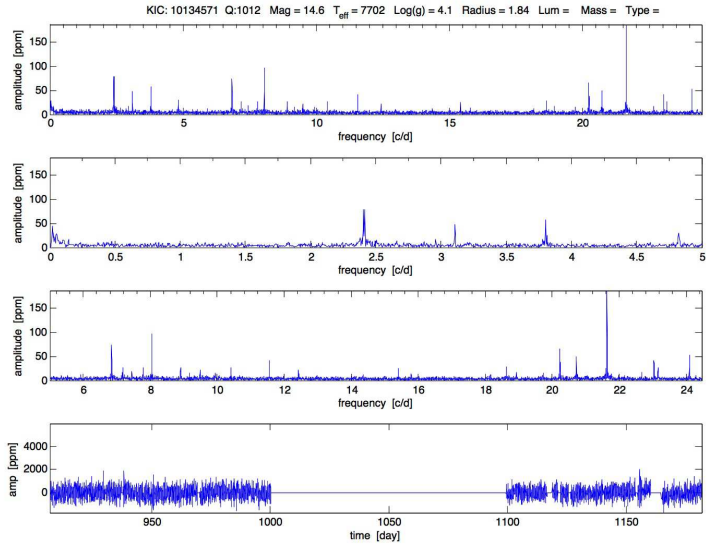}
\caption{KIC 10134571 light curve (bottom panel) and Fourier transform (top panels) showing many frequencies in both the $\gamma$ Dor and $\delta$ Sct range.  However, this star has T$_{\rm eff}$  = 7931 K (including the temperature offset from the KIC value), hotter than expected for a hybrid $\gamma$ Dor/$\delta$ Sct variable (see Fig. \ref{gammaDordeltaScutiLargerSym})}
\label{10134571}
\end{figure*}

\begin{figure*}%tb
\center
\includegraphics[width=1.5\columnwidth]{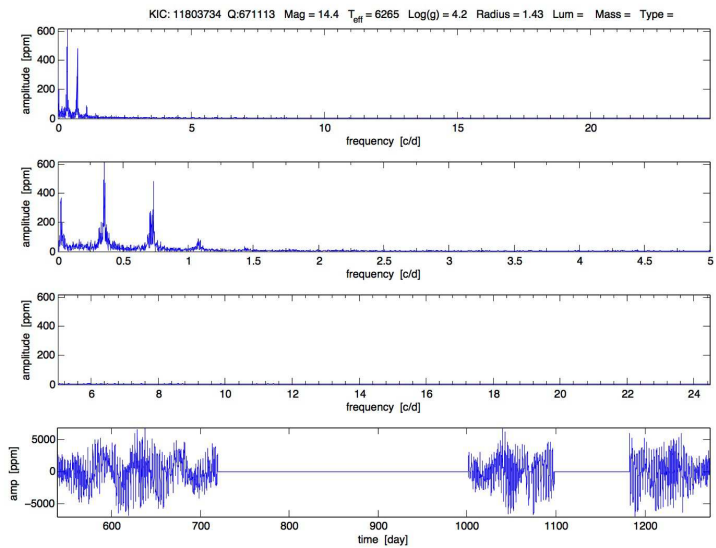}
\caption{KIC 11803734 light curve (bottom panel) and Fourier transform (top panels) showing peaks in the $\gamma$ Dor frequency range.  This star is the coolest $\gamma$ Dor candidate in our sample with T$_{\rm eff}$  = 6494 K (including the temperature offset from the KIC value), and is to the red of the $\gamma$ Dor instability region in Fig. \ref{gammaDordeltaScutiLargerSym}.}
\label{11803734}
\end{figure*}

\begin{figure*}%tb
\center
\includegraphics[width=1.5\columnwidth]{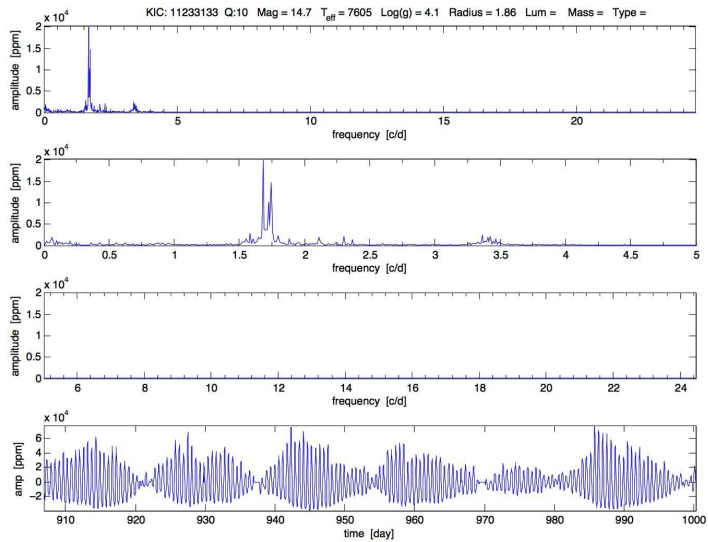}
\caption{KIC 11233133 light curve (bottom panel) and Fourier transform (top panels) showing peaks in the $\gamma$ Dor frequency range. This light curve and Fourier transform are typical of many $\gamma$ Dor candidates discovered in the {\it Kepler} data.  The light curve modulation in amplitude during periods of about 15 days is caused by the largest-amplitude modes with frequencies differing by about 1/15 cycles/day beating against each other, causing constructive or destructive interference in the light curve.  This star is the hottest $\gamma$ Dor candidate in our sample with T$_{\rm eff}$ = 7833 K (including the temperature offset from the KIC value), and is to the blue of the $\gamma$ Dor instability region in Fig. \ref{gammaDordeltaScutiLargerSym}.}
\label{11233133}
\end{figure*}

\begin{figure*}%tb
\center
\includegraphics[width=1.5\columnwidth]{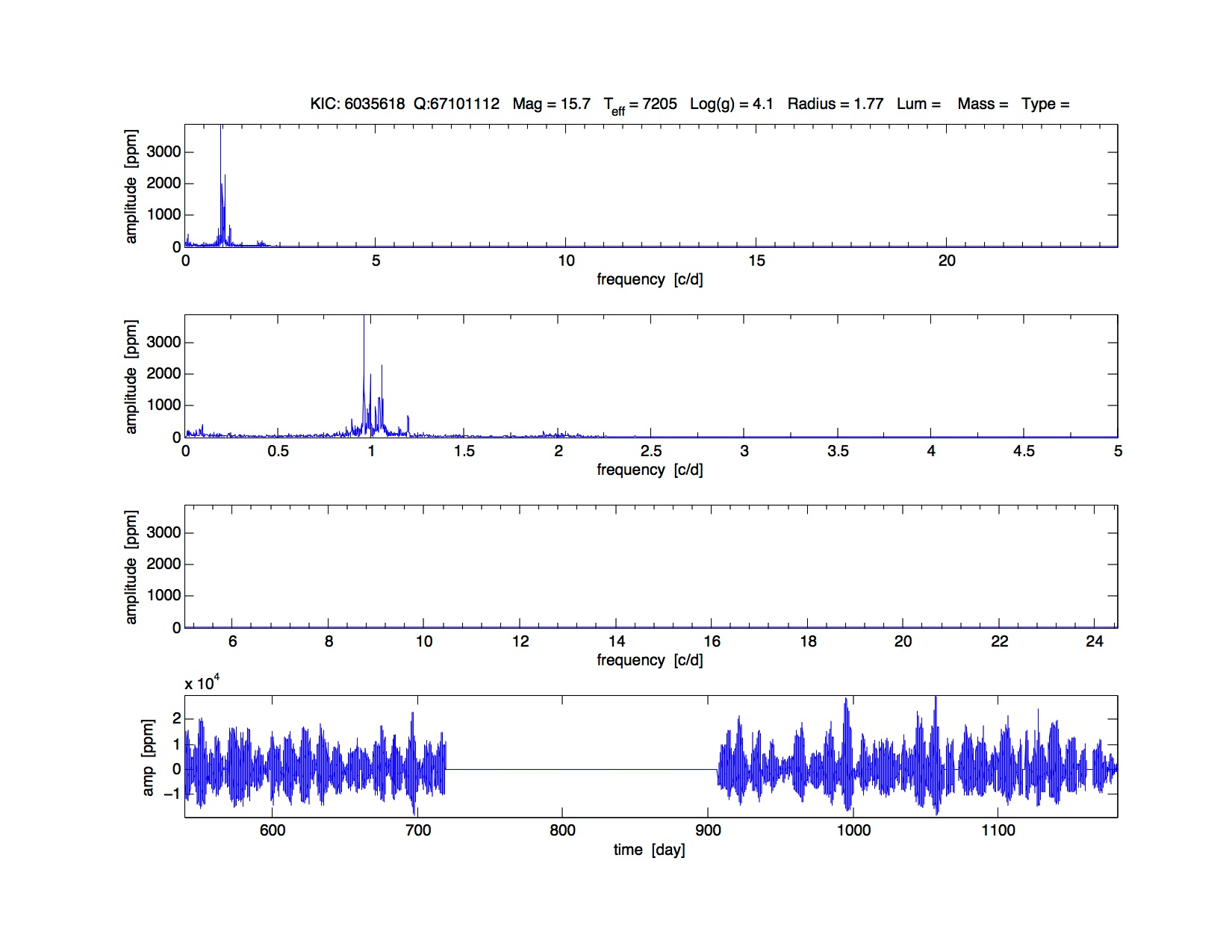}
\caption{KIC 6035618 light curve (bottom panel) and Fourier transform (top panels) showing peaks in the $\gamma$ Dor frequency range. This light curve and Fourier transform are typical of many $\gamma$ Dor candidates discovered in the {\it Kepler} data.  This star has T$_{\rm eff}$  = 7433 K (including the temperature offset from the KIC value), and is at blue edge of the $\gamma$ Dor instability region in Fig. \ref{gammaDordeltaScutiLargerSym}.}
\label{6035618}
\end{figure*}

\begin{figure*}%tb
\center
\includegraphics[width=1.5\columnwidth]{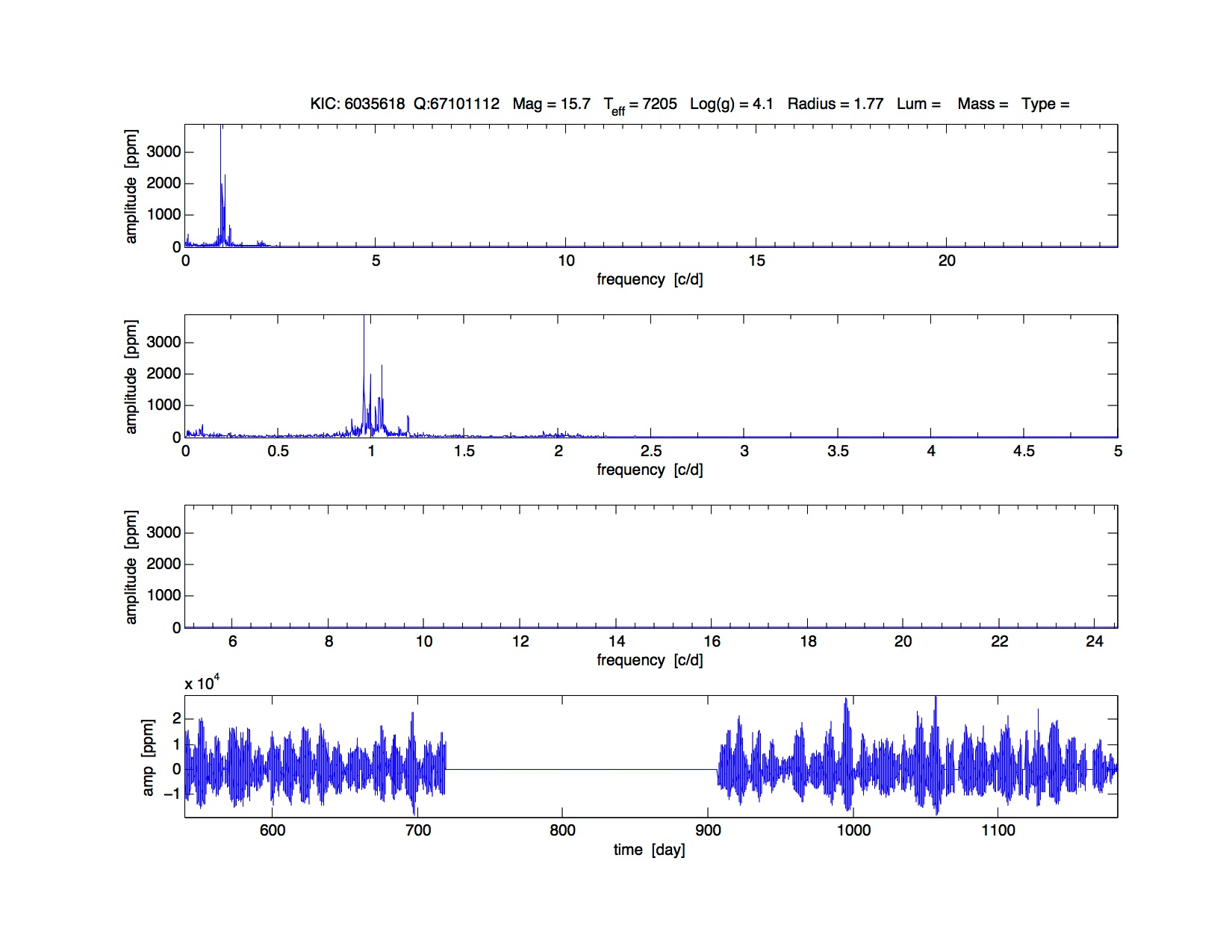}
\caption{KIC 7685157 light curve (bottom panel) and Fourier transform (top panels) showing peaks in the $\gamma$ Dor frequency range. The Fourier transform of this $\gamma$ Dor candidate also shows two regions at low frequency with high-amplitude peaks. This star has T$_{\rm eff}$   = 7297 K (including the temperature offset from the KIC value) and lies within the $\gamma$ Dor instability region in Fig. \ref{gammaDordeltaScutiLargerSym}.}
\label{7685157}
\end{figure*}

It is beyond the scope of this paper to consider and categorize the remaining `non-constant' stars. \citep[see][for more information on the variables discovered]{bradley14a, bradley14b}; however, we do want to show how these stars are distributed in relation to the instability regions established from ground-based observations.  In addition, we show a few examples of the light curves and location of $\delta$ Sct and $\gamma$ Dor candidates discovered in our sample to demonstrate that they can be identified readily from their light curves and Fourier transforms, and have frequencies with amplitude far above the noise level.  
We also want to point out that some of our  $\delta$ Sct and $\gamma$ Dor candidates lie within the expected instability regions, but some also lie also outside of them.

Figure \ref{NonconstantStarLargerSym} shows the distribution of the remaining 274 `non-constant' stars in our sample on the log g--T$_{\rm eff}$ diagram.  The light curves of most of these stars show irregular low-frequency variations consistent with rotating spots, stellar activity, or else are probable eclipsing binaries, so they are not all pulsating variables.

We examined the light curves and Fourier transforms of these Ônon-constantÕ stars, and identified 26 stars that appear unambiguously to be $\gamma$ Dor or $\delta$ Sct candidates. Figure \ref{gammaDordeltaScutiLargerSym} shows the location of these stars in the log g--T$_{\rm eff}$ diagram.   Some of these stars are much cooler than the $\gamma$ Dor instability strip red edge, even when uncertainties in T$_{\rm eff}$ and log g are taken into account.  Many $\gamma$ Dor and $\delta$ Sct candidates were also found outside of the instability-strip boundaries in the large sample of generally brighter {\it Kepler} stars by \citet{uytterhoeven11}.  Of these 26 pulsating star candidates, four are $\delta$ Sct candidates, 21 are $\gamma$ Dor candidates, and one is a hybrid candidate.

Figures \ref{7048016}, \ref{10134571}, \ref{11803734}, \ref{11233133}, \ref{6035618} and \ref{7685157} show light curves and Fourier transforms for six of these stars.  Some lie within the instability region in Fig. \ref{gammaDordeltaScutiLargerSym}, but are pulsating in unexpected frequencies, while some lie outside the instability regions.  Comments on the light curves and Fourier transforms are in the figure captions.  We note that sometimes, even for these faint 14-15.5 magnitude stars, when pulsations exist, they are readily visible in the {\it Kepler} data, with amplitudes of several parts per thousand or more, and dozens of frequencies visible in the Fourier transform.

\section{Conclusions and Future Work}

From our sample of 633 A-F main-sequence stars observed by {\it Kepler}, 359 stars, or roughly 60\%, were found to be `constant', defined as showing no frequencies above 20 ppm in their light curves between 0.2 and 24.4 c/d. 

Perhaps not surprisingly, we found very few (only six) photometrically non-varying stars within the $\gamma$ Dor and $\delta$ Sct pulsation instability regions.  However, the lack of variability in these several stars requires explanation.  

Of the 274 stars in our sample that show frequency peaks in their Fourier transform with amplitude greater than 20 ppm, 26 stars, or about 10\%, appear to be likely $\gamma$ Dor or $\delta$ Sct variable star candidates.  However, many of these stars do not lie within their expected instability regions, and their behavior also requires explanation.  

As a next step, we will extend our analysis to include many more stars observed during Quarters 14-17 (Cycle 4).  See \citet{guzik14} for preliminary Q14-16 results.

We would like to use stellar modeling to determine how much diffusive settling is possible and necessary to eliminate pulsations in $\gamma$ Dor or $\delta$ Sct stars.  We also hope to carry out model calculations to better determine theoretical instability-strip boundaries, and explore alternate pulsation driving mechanisms to help explain the {\it Kepler} stars that show pulsations at unexpected frequencies or lie outside of predicted instability boundaries.

\section*{Acknowlegements}
The authors are grateful for data and funding through the NASA {\it Kepler} Guest Observer Program Cycles 1-4.  KU acknowledges financial support by the Spanish National Plan of R\&D for 2010, project AYA2010-17803. This work has benefited from funding by the Project FP7-PEOPLE-IRSES:ASK no 269194.

\end{document}